# The development of low-temperature atomic layer deposition of $HfO_2$ for TEM sample preparation on soft photo-resist substrate

Kang-Ping Peng[1], Ya-Chi Liu[2], I-Feng Lin[3], Chih-Chien Lin[4], Shu-Wei Huang[5], and Chao-Cheng Ting*

[1]Department of Electronics Engineering, National Chiao Tung University, 1001 Ta Hsueh Road, Hsinchu, Taiwan 30010, ROC
[2]OptMic Lab, 340 S. Lemon Ave. #8365, Walnut, CA, 91798, USA
[3]Atom SemiconTek Co., Ltd, B35 1A2, No.1, Lising 1st Rd., East Dist., Hsinchu City 30078, Taiwan
[4]Center for Micro/Nano Science and Technology, National Cheng Kung University, 701, Tainan, Taiwan
[5]Department of Electrical, Computer and Energy Engineering, University of Colorado, Boulder, CO, USA
*Department of Materials Science and Engineering, National Chiao Tung University, 1001 Ta Hsueh Road, Hsinchu, Taiwan 30010, ROC
Phone: +86918528040 *Email: chaochengting.mse00g@g2.nctu.edu.tw

*Abstract-* **In this study, the method of low-temperature atomic layer deposition (ALD), which is applied on the soft photo-resist (PR) substrate forming hafnium dioxide ($HfO_2$) at 40°C to 85°C, is reported for the first time. This reveals the potential application in the TEM sample preparation. The thickness, refractive index, band gap, and depth profiling chemical state of the thin film are analyzed by ellipsometry, X-ray diffraction, and photoelectron spectroscopy respectively. Our TEM image shows a clear boundary between the photo-resist and hafnium dioxide deposited on PR, which indicates the low-temperature atomic layer deposition (ALD) may lead a new way for TEM sample preparation in advanced technology node.**

I. INTRODUCTION

In spite of high-energy sputtering during the Focus Ion Beam (FIB) process, FIB is commonly used in TEM sample preparation. To obtain a damage-free specimen, a protection layer on PR (photo-resist) substrate is required. The main deposition processes for the layer are RF sputtering Pt and PECVD oxide [1]. However, a soft PR layer has been developed in the advanced technology node to reduce the turbulence during etching process. This results in the current deposition processes for the protection layer not suitable for this application. With the use of RF sputtering thin film deposition method, the physical bombardment of RF sputtered metal particles alters the structure of the PR pattern. In the interim, the $O_2$ plasma induces radical based oxidation, which causes severe damage to the PR structure by PECVD. In addition, the high deposition temperature of CVD also limits its application in PR protection because the higher temperature will induce PR reflows. Atomic layer deposition technique is widely used in electronic industry [2]. However, few research addresses the challenges when the deposition is on the patterned soft PR substrate with the temperature is lower than 100°C [3]. In this research, the method of low-temperature thermal atomic layer deposition (ALD) to form $HfO_2$ at 40°C to 85°C is developed. The preposition is preformed on the soft and patterned PR substrate with TEM examination the step coverage of the ALD thin film on the patterned soft PR substrate for the first time. With the use of ellipsometry, the refractive index and energy band gap show no significantly difference in the thin film quality within the deposition temperatures. The detailed physics of converting the refractive index and energy band gap is described in the work of Selj, J. H., et al.[4]. The carbon residue of interface from the condensation of metal organic precursor is discovered in the XPS depth profile. In addition, a slightly cross diffusion photo electron signal of carbon 1s and Hf $4f_{7/2}$ is also noticed. This might be induced by the diffusion of Hf precursor to the affluent of organic solvent in soft PR. A severe particle generation is observed when the deposition temperature is below 40°C. The TEM cross section image shows a clear boundary between patterned PR and the low temperature ALD $HfO_2$ thin film. This is also shown in EDX mapping image. The unwanted particle generation may be caused by the incomplete ALD reaction process. For 40°C-85°C, there is no unwanted particle generation based on our proposed method, which indicates the potential application of low-temperature ALD in TEM sample preparation on the soft photo-resist substrate in advanced technology node is promising.

II. EXPERIMENT DETAILS

The experiment is arranged on the PR coated wafer with thickness of 50nm. The PR (INTERVVIA 8023) is commercially obtained from DOW chemical. The PR is only conducted with a soft bake process in the condition of 100°C for 10 minutes before the atomic layer deposition (ALD). We create the patterned PR substrate by using focused electron beam bombardment on the PR with a step structure for further step coverage examination. A 20nm ALD hafnium dioxide is deposited via ATOM FA-200 system with metal organic compound as the precursor, and Ar as the carrier gas. Each ALD reaction cycle consists of four steps. The PR substrate is exposed to a Hf precursor pulse of 6s with the Ar carrier gas flow of 30sccm, followed by an Ar purge of 200sccm for 180s. The difference in the thin film quality at the deposition temperatures ranging from 40°C to 85°C is studied via ellipsometry, X-ray diffraction, and photoelectron spectroscopy (XPS, Thermo VG 350) using a Mg Kα X-ray source. The XPS depth profile is using for the analysis of the boundary of multi-film structures. The signal of carbon is for binding energy calibration. FIB and TEM are performed to evaluate the boundary between the soft patterned photo resist and the ALD $HfO_2$. 150 ALD cycles are performed at different temperatures. The ALD deposition parameters are listed in Table I.

## III. RESULTS AND DISCUSSION

Fig. 1 shows the stacked thin-film schematic of $HfO_2$ deposited on photo-resist contains substrate. The thickness of PR(50nm) is adjusted by the rotation speed. The 10 minutes soft bake is removing as much unwanted solvent as possible, but still retaining the similar soft surface as the challenge we are facing in the advanced technology node. The $HfO_2$ is deposited with 150 ALD cycles at 40°C-85°C. The thickness is ranging from 20.7nm to 26.2nm at 40°C - 85°C respectively.

Table I. The list of ALD deposition parameters

| Deposition parameters | Setting |
|---|---|
| Substrate temperature | 40-85 °C |
| Base pressure | 1×10⁻³ torr |
| Precursor pulse | Ar 30sccm/6s |
| Precursor Purge | Ar 200sccm/180s |
| Deposition pressure | 1×10⁻² torr |
| $H_2O$ gas flow | 50sccm/1s |
| Final Purge | Ar 200sccm/180s |

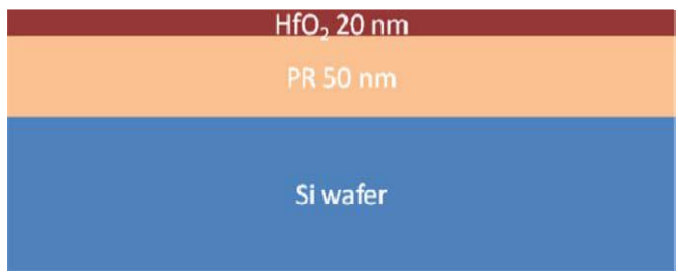

Fig. 1 The schematic diagram of $HfO_2$ thin film deposited on photo-resist contains substrate.

Fig. 2 shows the XPS depth profile of the specimen at the deposition temperature of 85°C. The Hf4f XPS signal reveals that the Hf atom peak is at the depth of 20nm, then dipping sharply as tailing to the depth of 30nm. We also discover an interesting C1s signal at the depth between 20 and 25nm. This signal was suspected not from the same origin of the C1s signal at the depth between 30 and 80nm. This signal (20-25nm) is belonged to the carbon atom from the condensation of the metal organic precursor due to the overlapping with the Hf signal in the same depth region.

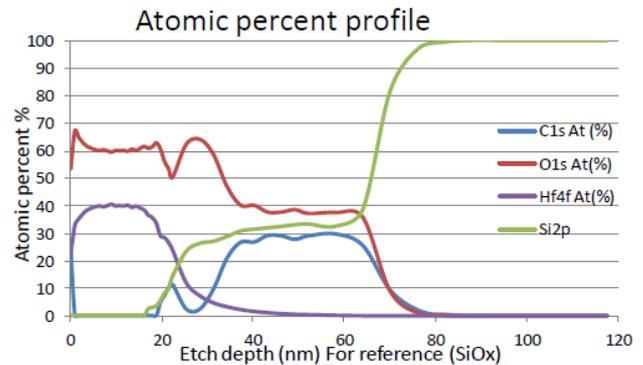

Fig. 2 The XPS depth profile of the specimen at the deposition temperature of 85°C

Fig. 3 is the XRR plot of the $HfO_2$ at 50°C. The data shows that the thickness of $HfO_2$ is decreased as the deposition temperature increased, which is aligned the result with ellipsometry.

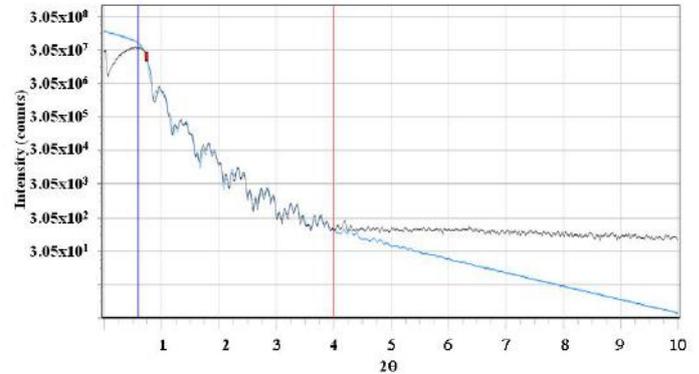

Fig. 3 XRR plot of PEALD 150 cycles at the deposition temperature of 50°C.

Fig. 4 is the plot of thickness, refractive index, and band gap of $HfO_2$ with 150 ALD cycles at the temperature of 40-85°C. The value of refractive index is ranging from 1.93 to 1.89 at the deposition temperature of 40 °C to 85°C, which variation rate is 2% in total. This indicates that quality of the $HfO_2$ thin film does not alter as the deposition temperature decreases. However, the variation rate in thickness of the thin film is larger than it in the value of refractive index, and band gap. The largest thickness is shown at 40°C, in the lowest deposition temperature of the ALD condition. This may be a result of CVD reaction mechanism from the excess precursor residue on the PR surface, and cannot be removed easily during the purge process at the lower temperature. The CVD phenomenon is even more obvious when the deposition temperature is lower than 40ºC. In spite of increasing purge time, a severe particle generation occurs as the temperature is lower than 40ºC. The detail of the ALD thin film property at temperature of 40-85ºC with 150 ALD cycles are listed in table II.)

Table II. The list of low-temperature ALD thin film property at temperature of 40-85ºC with 150 ALD cycles

| ALD temp (ºC) | Thickness (A) | n @ 633nm | Band Gap (eV) |
|---|---|---|---|
| 40 | 262 | 1.9355 | 6.54 |
| 50 | 248 | 1.934 | 6.46 |
| 70 | 220 | 1.9195 | 6.33 |
| 85 | 207 | 1.8928 | 6.26 |

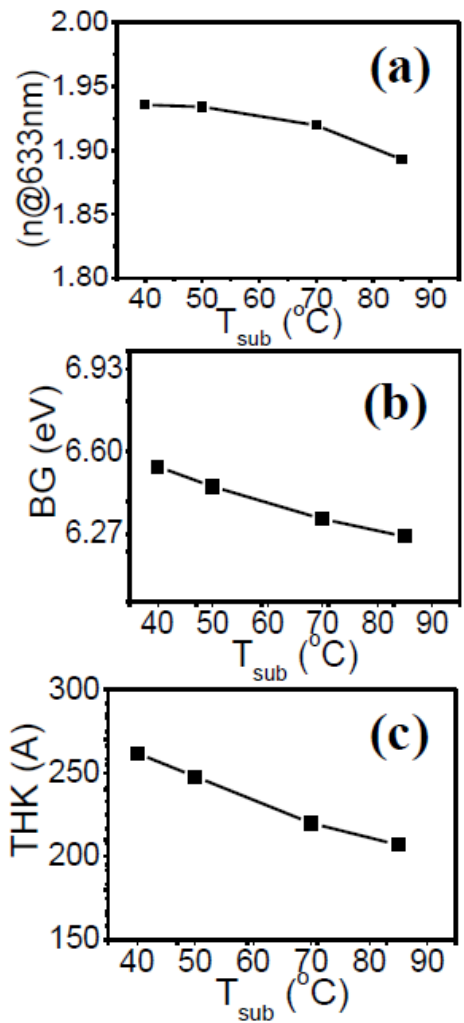

Fig. 4 (a) thickness, (b) refractive index, and (c) band gap plot of $HfO_2$ with 150 ALD cycles at the temperature of 40 to 85ºC.

Fig. 5(a) shows the cross section of $HfO_2$ thin-film with 150 ALD cycles on the planner photo-resist contained substrate. The composite EDX element mapping of the TEM sample is presented in fig. 5 (b). It can be easily discovered the interface between $HfO_2$ and the PR under layer is clear.

Fig. 6(a) shows The TEM cross section image of $HfO_2$ thin film with 150 ALD cycles at 85ºC on patterned photo-resist substrate. The ALD thin-film are well covered with excellent step coverage even on the small particle located on the left bottom in the image. It is worth noticing that the ALD thin-film still demonstrates excellent step coverage with the landing angle <90º illustrates in the fig. 6 (b) the high magnitude TEM image. Fig. 6(c,e) is the Hf, oxygen EDX element mapping of the specimen indicates a continuous ALD $HfO_2$ thin-film are developed. We also discover the void occurred in the PR thin film which is caused by the prolonged electron beam irradiation during the TEM inspection. The image is presented in Fig. 6(d).

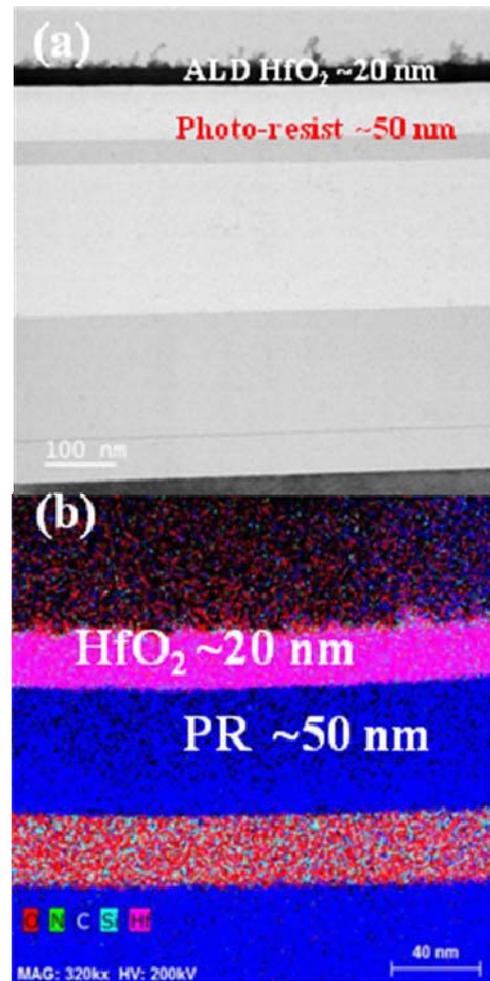

Fig. 5 (a) The TEM cross section image of $HfO_2$ thin film with 150 ALD cycles on photo-resist contained substrate. (b) The composite EDX element mapping of the TEM sample.

However, no deformation or distortions are discovered base on the flat and straight surface above the patterned PR substrate shows in fig. 6(a). These findings show ALD is a promising technique for TEM sample preparation on patterned PR substrate.

## IV. CONCLUSION

In this study, a low-temperature, 40°C to 85°C, ALD of $HfO_2$ technique for TEM sample preparation on soft patterned PR substrate is successfully developed. This has been the first time so far to demonstrate the feasibility of ALD $HfO_2$ on soft patterned photo-resist substrate. The TEM image of flat PR substrate reveals clear boundary between the PR and $HfO_2$. The excellent step coverage is discovered on the patterned photo-resist substrate even with landing angle < 90°. The ALD thin-film structure remains intact on the void PR substrate during the electron beam irradiation. Our findings demonstrate ALD is crucial for the TEM sample preparation in the development of advanced technology node.

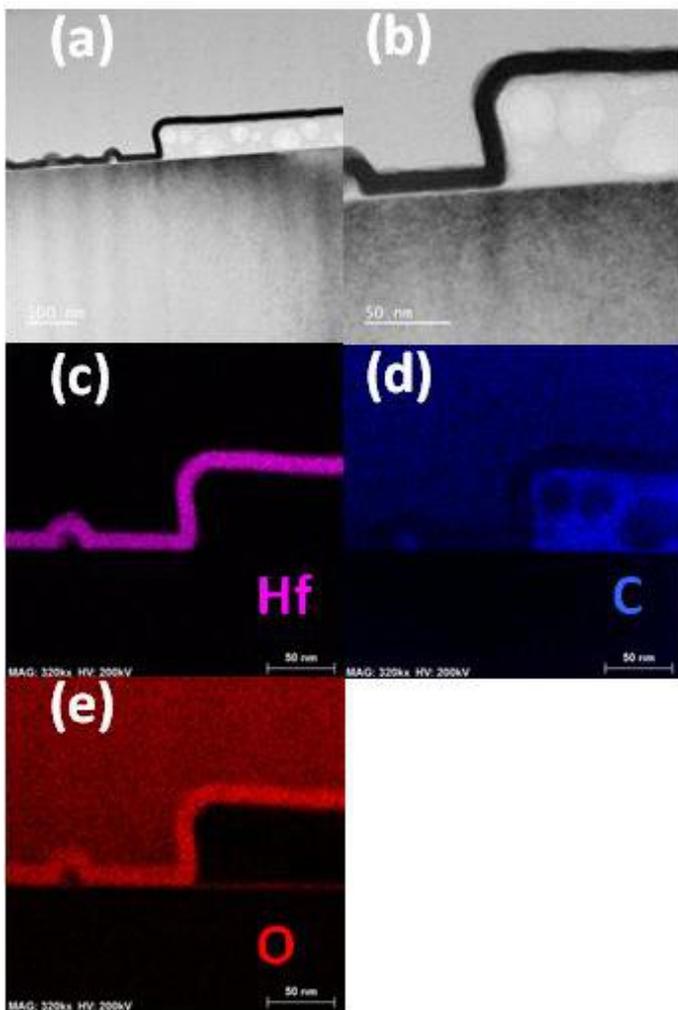

Fig. 6 (a) The TEM cross section image of $HfO_2$ thin-film with 150 ALD cycles on patterned photo-resist substrate at 85°C. (b) The high magnitude TEM image. (c) The Hf EDX element mapping of the TEM sample. (d) The carbon EDX element mapping of the tem sample. (e) The oxygen EDX element mapping of the tem sample.

## ACKNOWLEDGMENT

The authors acknowledge the support of the Ministry of Science and Technology, Taiwan.